\documentclass[a4paper,11pt]{article}
\pdfoutput=1
\usepackage{jheppub}

\usepackage[inline]{enumitem}
\usepackage{amsmath}
\usepackage{hyperref,url}
\usepackage[dvipsnames]{xcolor}
\usepackage{graphicx,tabularx,tabularray}
\usepackage{enumitem,color,xcolor}
\usepackage{natbib}
\usepackage{cancel}
\usepackage[compat=1.1.0]{tikz-feynman}
\usepackage{float}

\definecolor{dgreen}{rgb}{.0,.6,.0}
\definecolor{lime}{HTML}{A6CE39}
\definecolor{lg}{RGB}{220,220,220}

\DeclareRobustCommand{\orcidicon}{\hspace{-2.1mm}
\begin{tikzpicture}
\draw[lime,fill=lime] (0,0.0) circle [radius=0.13] node[white] {{\fontfamily{qag}\selectfont \tiny ID}}; \draw[white,fill=white] (-0.0525,0.095) circle [radius=0.007]; 
\end{tikzpicture} \hspace{-3.7mm} }
\foreach \x in {A, ..., Z} {\expandafter\xdef\csname orcid\x\endcsname{\noexpand\href{https://orcid.org/\csname orcidauthor\x\endcsname} {\noexpand\orcidicon}}}

\newcommand{\Eprint}[1]{\href{#1}}

\preprint{PSI-PR-23-30, ZU-TH 46/23, ICPP-72}

\title{ \boldmath Uncovering New Higgses in the LHC Analyses of Differential $t\bar t$ Cross Sections}

\author[a,b]{Sumit Banik}
\author[a,b]{Guglielmo Coloretti\orcidA{}}
\author[a,b]{Andreas Crivellin\orcidB{}}
\author[c,d]{Bruce Mellado}
\affiliation[a]{Physik-Institut, Universit\"at Z\"urich, Winterthurerstrasse 190, CH--8057 Z\"urich, Switzerland}
\affiliation[b]{Paul Scherrer Institut, CH--5232 Villigen PSI, Switzerland}
\affiliation[c]{School of Physics and Institute for Collider Particle Physics, University of the Witwatersrand,
Johannesburg, Wits 2050, South Africa}
\affiliation[d]{iThemba LABS, National Research Foundation, PO Box 722, Somerset West 7129, South Africa}

\emailAdd{sumit.banik@psi.ch}
\emailAdd{guglielmo.coloretti@physik.uzh.ch}
\emailAdd{andreas.crivellin@psi.ch}
\emailAdd{bmellado@mail.cern.ch}

\abstract{ 
Statistically significant tensions between the Standard Model (SM) predictions and the measured lepton distributions in differential top cross-sections emerged in LHC Run~1 data and became even more pronounced in Run~2 analyses. Due to the level of sophistication of the SM predictions and the performance of the ATLAS and CMS detectors, this is very remarkable. Therefore, one should seriously consider the possibility that these measurements are contaminated by beyond-the-SM contributions. In this article, we use the differential lepton distributions from the latest ATLAS $t\bar t$ analysis to study a new physics benchmark model motivated by existing indications for new Higgses: a new scalar $H$ is produced via gluon fusion and decays to $S^\prime$ ($95\,$GeV) and $S$ ($152\,$GeV), which subsequently decay to $b\bar b$ and $WW$, respectively. In this setup, the total $\chi^2$ is reduced, compared to the SM, resulting in $\Delta\chi^2=34$ to $\Delta\chi^2=158$, corresponding to a significance of $5.8\sigma$ to $13\sigma$, depending on the SM simulation used. Notably, allowing $m_S$ to vary, the combination of the distributions points towards $m_S\!\approx\!150\,$GeV, which is consistent with the existing $\gamma \gamma$ and $WW$ signals, rendering a mismodelling of the SM unlikely. Averaging the results of the different SM predictions, $\sigma(pp\to H\to SS^\prime)\times$Br$(S\to WW)\times$Br$(S^\prime\to bb)\approx9$pb is preferred. Assuming that $S^\prime$ is SM-like, the $95\,$GeV $\gamma\gamma$ excess can be explained if $S$ decays dominantly to $W$ bosons. That latter suggests that $S$ is the neutral component of the $SU(2)_L$ triplet with hypercharge~0. 
}
\newpage
\begin{document}
\maketitle

\section{Introduction} 
The known fundamental constituents of matter and their interactions (excluding gravity) are described by the Standard Model (SM) of particle physics. It has been successfully tested and verified by a plethora of measurements~\cite{ParticleDataGroup:2020ssz} with the discovery of the Brout-Englert-Higgs boson~\cite{Higgs:1964ia,Englert:1964et,Higgs:1964pj,Guralnik:1964eu} at the LHC~\cite{Aad:2012tfa,Chatrchyan:2012ufa} providing its last missing ingredient. Furthermore, this 125$\,$GeV boson has properties and decay rates~\cite{Langford:2021osp,ATLAS:2021vrm} in agreement with the SM expectations.

Nonetheless, a multitude of new physics models predicts the existence of new Higgses (i.e.~scalars that are uncharged under the strong force). Such models are viable if the SM Higgs signal strengths are not significantly altered and their contribution to the $\rho$ parameter ($\rho=m_Z^2/(m_W^2/\cos^2\theta_W)$) is sufficiently small (i.e.~custodial symmetry is approximately respected). In fact, the global average of the $W$-boson mass measurements~\cite{ALEPH:2013dgf,LHCb:2021bjt,CDF:2022hxs,ATLAS:2023fsi} even points to a small additional violation of custodial symmetry with a significance of $3.5\sigma$~\cite{LHC-TeVMWWorkingGroup:2023zkn}. This can be explained by various extensions of the SM scalar sector~\cite{Strumia:2022qkt}. In particular, the $SU(2)_L$ triplet with hypercharge 0 is an interesting option for explaining the $W$ mass since it predicts a necessarily positive shift, in agreement data.

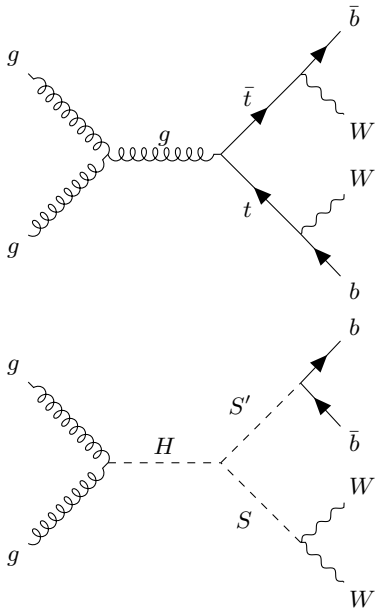
\begin{figure}[h!]
    \centering
    \begin{tikzpicture}[baseline=(current bounding box.center)]
            \begin{feynman}
            \vertex (a);
            \vertex [above left=1.5cm of a] (c) {$g$};
            \vertex [below left=1.5cm of a] (d) {$g$};
            \vertex [right=1.5cm of a] (b) ;
            \vertex [above right=1.5cm of b] (e);
            \vertex [below right=1.5cm of b] (f);
            
            \vertex [above right=0.75cm of e] (i) {$\bar b$};
            \vertex [below right=0.75cm of e] (j) {$W$};
            \vertex [above right=0.75cm of f] (k) {$W$};
            \vertex [below right=0.75cm of f] (l) {$b$};
            \diagram{
                (d) -- [gluon] (a) -- [gluon] (c);
                (a) -- [gluon, edge label=$g$] (b);
                (f) -- [fermion, edge label=$t$] 
                (b) -- [fermion, edge label=$\bar t$] (e);
                (j) -- [boson] (e) -- [fermion] (i);
                (l) -- [fermion] (f) -- [boson] (k);
            };
        \end{feynman}
            \end{tikzpicture} \hspace{2cm}
                \begin{tikzpicture}[baseline=(current bounding box.center)]
        \begin{feynman}
            \vertex (a);
            \vertex [above left=1.5cm of a] (c) {$g$};
            \vertex [below left=1.5cm of a] (d) {$g$};
            \vertex [right=1.5cm of a] (b) ;
            \vertex [above right=1.5cm of b] (e);
            \vertex [below right=1.5cm of b] (f);
            
            \vertex [above right=0.75cm of e] (i) {$b$};
            \vertex [below right=0.75cm of e] (j) {$\bar{b}$};
            \vertex [above right=0.75cm of f] (k) {$W$};
            \vertex [below right=0.75cm of f] (l) {$W$};
            \diagram{
                (d) -- [gluon] (a) -- [gluon] (c);
                (a) -- [scalar, edge label=$H$] (b);
                (f) -- [scalar, edge label=$S$] 
                (b) -- [scalar, edge label=$S^\prime$] (e);
                (j) -- [fermion] (e) -- [fermion] (i);
                (l) -- [boson] (f) -- [boson] (k);
            };
        \end{feynman}
    \end{tikzpicture}
    \caption{Feynman diagrams showing the leading SM contribution to top pair production and decay as well as the NP signal in our benchmark model contaminating the measurement of $t\bar t$ differential distributions.}
    \label{fig:feynman}
\end{figure}

Dedicated searches for additional Higgs bosons at the LHC have been mostly performed inclusively or with a limited number of topologies, such that significant regions of the phase space remain unexplored. In particular, associated production received relatively little attention. Nonetheless, indications for new Higgses at the electroweak (EW) scale with masses around 95$\,$GeV~\cite{LEPWorkingGroupforHiggsbosonsearches:2003ing,CMS:2018cyk,CMS:2022rbd,CMS:2022tgk,ATLAS:2023jzc} and 152$\,$GeV~\cite{ATLAS:2021jbf} arose, with global significances of $3.8\sigma$ and $4.9\sigma$~\cite{Crivellin:2021ubm,Bhattacharya:2023lmu}, respectively. In particular, the existence of a 152$\,$GeV boson is also motivated by the ``multi-lepton anomalies''~\cite{Buddenbrock:2019tua,vonBuddenbrock:2020ter,Hernandez:2019geu,Fischer:2021sqw}. These are processes involving multiple leptons and missing energy, with and without ($b$-)jets, where deviations from the SM expectations have been observed over the last years. Processes with such signatures (in the SM) include $WW$, $WWW$, $Wh$, $tW$, $t\bar t$, $t\bar tW$ and $t\bar t t\bar t$ (see Ref.~\cite{Fischer:2021sqw} and references therein). In particular, $WW$ signals are compatible with, or even suggest, a mass of $\approx\!150\,$GeV~\cite{vonBuddenbrock:2017gvy,Coloretti:2023wng}.

Here, we will study the statistically most significant multi-lepton excess encoded in the latest ATLAS analysis of the $t\bar t$ differential cross-sections~\cite{ATLAS:2023gsl}. Differential lepton distributions (from leptonic $W$ decays) are advantageous in this context, since the total $t\bar t$ cross-section is large and very sensitive to QCD corrections~\cite{Beneke:2011mq,Barnreuther:2012wtj,Czakon:2012zr,Kidonakis:2023jpj}. Such differential measurements have been performed by ATLAS~\cite{ATLAS:2017dhr,ATLAS:2019hau,ATLAS:2023gsl} and CMS~\cite{CMS:2018adi,CMS:2020djy}. The ATLAS analysis~\cite{ATLAS:2023gsl} of $t\bar t$ does not only provide the invariant di-lepton mass ($m^{e\mu}$) and the angle between the leptons ($\Delta \phi^{e\mu}$), but also contains an extensive analysis of the different SM predictions using various combinations of Monte Carlo (MC) simulators.\footnote{Note that the CMS analysis of $t+W$~\cite{CMS:2018amb} is less precise but consistent with the ATLAS findings~\cite{Buddenbrock:2019tua}. Furthermore, note that for the $m_{t \bar t}$ distribution, we need the $W$ bosons to decay hadronically, which is, however, much less precise than the leptonic distributions.} Importantly, Ref.~\cite{ATLAS:2023gsl} concluded that: ``No model (SM simulation) can describe all measured distributions within their uncertainties.'' Due to the level of rigour and sophistication of the LHC simulations~\cite{ATLAS:2010arf,CMS:2019csb} and the performance of the ATLAS and CMS detectors~\cite{ATLAS:2019qmc,CMS:2015xaf,CMS:2012nsv}, this very significant disagreement in a high-statistics measurement including leptons is remarkable.

\begin{figure*}[t!]
    \includegraphics[width=0.78\linewidth]{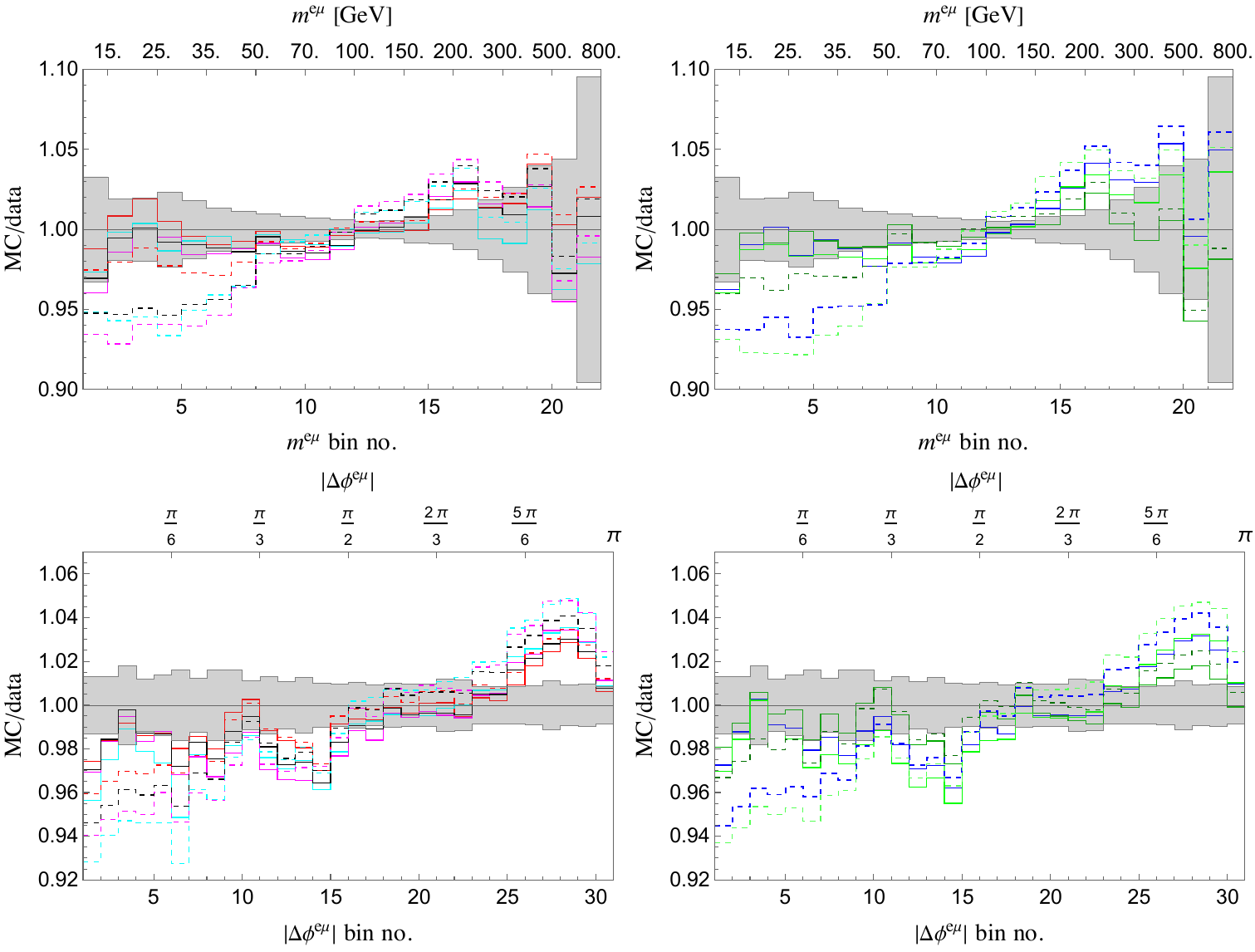}
    \raisebox{0.26\height}{\includegraphics[width=0.21\linewidth]{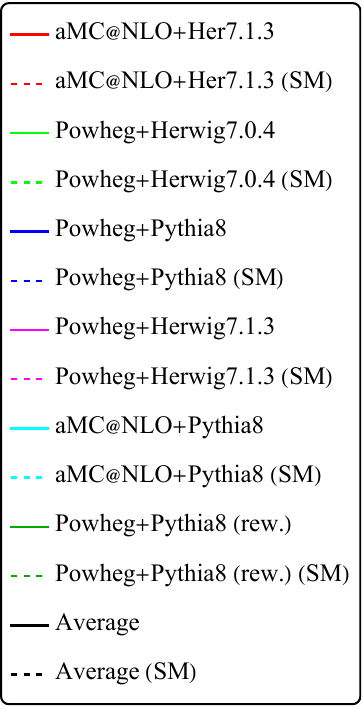}}
    \caption{The dashed colored lines show six different SM predictions (MC) normalized to data given in Ref.~\cite{ATLAS:2023gsl}. The solid lines include the NP contribution from our benchmark model obtained by a combined global fit to $m^{e\mu}$ and $\Delta\phi^{e\mu}$ data. The black lines are obtained by averaging the six predictions and the grey band shows the total uncertainty (systematic and statistical). One can see that the agreement between theory and experiment is significantly increased by adding a NP effect.}
    \label{fig:mlldata}
\end{figure*}

Therefore, the possibility that NP contaminates the measurement of this SM process should be considered seriously. For the decay $t\to W b$ with $W\to \ell\nu$, the experimental signature of top pair production is opposite-sign different-flavour (to reduce the $Z,\gamma$ induced background) di-leptons with one or more $b$-jets. Because the deviations from the SM predictions are most pronounced at low invariant masses of the $e$--$\mu$ system ($m^{e\mu}$) and small transverse momenta, this points towards an electroweak (EW) scale extension of the SM. For such a beyond-the-SM explanation, light new particles are needed, which are both the source of bottom quarks and of opposite-sign different flavour di-leptons (or are at least produced in association with them). Since the excess is not localized in $m^{e\mu}$, this excludes the direct decay of a new particle to $e\mu$. Furthermore, the broad excess in $WW$ at low masses~\cite{CMS:2022uhn,ATLAS:2022ooq,Coloretti:2023wng} suggests a new scalar decaying to $W$ bosons. Finally, since a SM-like Higgs with a mass below $\approx100\,$GeV naturally decays dominantly to bottom quarks, this hints towards the decay chain $pp\to H\to SS^\prime$ with $S\to WW$ and $S^\prime \to b\bar b$ with $m_{S^\prime}=95\,$GeV and $m_S=152\,$GeV, motivated by the respective hints for narrow resonances.\footnote{Note that this parameter space is not covered by non-resonant Higgs pair analyses (which aim at higher transverse momenta) or searches for supersymmetric particles like sleptons or charginos (which require more missing energy).} The Feynman diagram giving the leading contribution in the SM and the one generating the potential NP contamination of our simplified NP model are shown in Fig.~\ref{fig:feynman}.


\begin{table}[!t]
\begin{center}
\resizebox{\textwidth}{!}{%
\begin{tblr}{colspec={Q[c,4.2cm]|Q[c,.5cm]Q[c,.5cm]Q[c,.5cm]Q[c,.9cm]|Q[c,.5cm]Q[c,.5cm]Q[c,.5cm]Q[c,.9cm]|Q[c,.5cm]Q[c,.5cm]Q[c,.5cm]Q[c,.6cm]Q[c,1.3cm]}, vline{3-5} = {2}{dashed}, vline{7-9} = {2}{dashed}, vline{11-14} = {2}{dashed}}

& \SetCell[c=4]{c} ${m^{e\mu }}$ & & & & \SetCell[c=4]{c} $\Delta {\phi ^{e\mu }}$ & & & & \SetCell[c=4]{c} ${m^{e\mu }} + \Delta {\phi ^{e\mu }}$\\

& $\chi _{{\rm{SM}}}^2$ & $\chi _{{\rm{NP}}}^2$ & $\sigma_{{\rm{NP}}}$ & Sig. & $\chi _{{\rm{SM}}}^2$ & $\chi _{{\rm{NP}}}^2$ & $\sigma _{{\rm{NP}}}$ & Sig. & $\chi _{{\rm{SM}}}^2$ & $\chi _{{\rm{NP}}}^2$ & $\sigma_{{\rm{NP}}}$ & Sig. & $m_S$[GeV] \\
\hline 
\textcolor{blue}{{\rm{Powheg+Pythia8}} } 
& 146 & 50 & 10pb & 9.8$\sigma$ & 183 & 73 & 11pb & 10.5$\sigma$ & 213 & 102 & 9pb & 10.5$\sigma$ & 143-156\\ 

\textcolor{red}{{{\rm{aMC@NLO+Herwig7}}{\rm{.1}}{\rm{.3}}}} 
& 31 & 13 & 4pb & 4.2$\sigma$ & 96 & 38 & 8pb & 7.6$\sigma$ & 102 & 68 & 5pb & 5.8$\sigma$ & $--$ \\
 
\textcolor{cyan}{{\rm{aMC@NLO+Pythia8}}} & 89 & 14 & 9pb & 8.7$\sigma$ & 277 & 83 & 15pb & 14.0$\sigma$ & 291 & 163 & 10pb & 11.3$\sigma$ & 148-157 \\

\textcolor{magenta}{{{\rm{Powheg+Herwig7}}{\rm{.1}}{\rm{.3}}}} & 138 & 32 & 10pb & 10.3$\sigma$ & 245 & 93 & 13pb & 12.3$\sigma$ & 261 & 126 & 10pb & 11.6$\sigma$ & 149-156\\

\textcolor{dgreen}{{\rm{Powheg+Pythia8~(rew)}}} & 40 & 12 & 5pb & 5.3$\sigma$ & 54 & 26 & 6pb & 5.3$\sigma$ & 69 & 35 & 5pb & 5.8$\sigma$ & $--$ \\
 
\textcolor{green}{{{\rm{Powheg+Herwig7}}{\rm{.0}}{\rm{.4}}}} & 186 & 41 & 12pb & 12.0$\sigma$ & 263 & 99 & 14pb & 12.8$\sigma$ & 294 & 126 & 12pb & 13.0$\sigma$ & 149-156\\

\hline
Average & 93 & 23 & 8pb & 8.4$\sigma$ & 172 & 63 & 11pb  & 10.4$\sigma$ & 182 & 88 & 9pb & 9.6$\sigma$ & 143-157\\
\end{tblr}}
\caption{$\chi^2$ values, preferred cross-section ($\sigma_{\rm NP}$) significance (Sig.) etc. for the individual $m^{e\mu}$ and $|\Delta\phi^{e\mu}|$ distributions and the combined fit to them for the six different SM simulations and their average. The $\chi^2_{\rm NP}$ is for our benchmark scenario with $m_S\approx152\,$GeV while the $m_S$ gives the preferred rage of it from the fit, assuming it to be a free parameter. A dash in the $m_S$ column means that the preferred $1\sigma$ rage is wider than $140\,$GeV--$160\,$GeV. Note that the cross section preferred by $m^{e\mu}$ and $|\Delta\phi^{e\mu}|$ individually are in very good agreement with each other. The preferred NP cross-section $\sigma_{\rm NP}$ is given for Br$[S^\prime\to b\bar b]\approx100\%$ and Br$[S\to WW]\approx100\%$.}
\label{tab:resmass150}
\end{center}
\end{table}

\section{Analysis and Results}
\label{sec:simplified}

ATLAS~\cite{ATLAS:2023gsl} considered six different SM simulations as given in Table~\ref{tab:resmass150}. In these setups, they predicted differential $t\bar t$ distributions, including the invariant mass of the electron-muon pair ($m^{e\mu}$), the sum of the energies $E^e+E^\mu$, the angle between the leptons ($|\Delta\phi^{e\mu}|$), the sum of transverse momentum of the leptons ($p_T^e+p_T^\mu$) and the pseudo-rapidity of the lepton system ($\eta$). Here, we will focus on $m^{e\mu}$ and $|\Delta\phi^{e\mu}|$, as they have a fine binning and show the most significant deviations from the SM predictions.\footnote{The high mass region for $m^{e\mu}$ could in general be relevantly affected by EW corrections. However, Ref.~\cite{Jezo:2023rht} found that at low masses, where the discrepancies are most pronounced, their impact is small. We check that excluding the high-mass bins from the fit would have a very small impact.} However, we will later predict the other different distributions to show the consistency of the NP explanation.

The main input for our analysis is thus Fig.~10 and 11 in Ref.~\cite{ATLAS:2023gsl}, showing the ratio of expected events within the SM (MC) over measured events (data) per bin $i$, ${\rm MC}|_{i}/{\rm data}|_{i}=x_i$, as well as the corresponding Tables~21 and 24 where the normalized cross sections and the (relative) systematic and statistical uncertainties for each bin ($\delta_i$) are given. Note that ATLAS normalized the differential cross-section to the total one such that the resulting $\chi^2$ value is minimized. In particular, $m^{e\mu}$ is given in 21 bins between \{0.0, 15.0, 20.0, 25.0, 30.0, 35.0, 40.0, 50.0, 60.0, 70.0, 85.0, 100.0, 120.0, 150.0, 175.0, 200.0,  250.0, 300.0, 400.0, 500.0, 650.0, 800.0\}\,GeV. As the binning for low values of $m_{e\mu}$ is relatively thin, we display the data in Fig.~\ref{fig:mlldata} with equal size for all bins to give also optically equal weight to each of them. For $|\Delta\phi^{e\mu}|$ each bin corresponds to an angle of $\pi/30$ and we therefore have 30 bins.

In order to obtain the SM prediction for each bin (and thus $x_i$) for the different distributions, ATLAS generated $t\bar t$ samples with several different matrix element generators, parton shower, and fragmentation simulations. In particular, the nominal sample was obtained using  {\tt Powheg~Box}~\cite{Frixione:2007vw,Alioli:2010xd,Frixione:2007nw} with {\tt NNPDF3.0~PDF}~\cite{NNPDF:2014otw} interfaced with {\tt Pythia~8.230}~\cite{Sjostrand:2006za,Sjostrand:2014zea}. Alternative simulations used {\tt MADGRAPH\_AMC$@$NLO}~\cite{Alwall:2014hca} ({\tt aMC$@$NLO}) with {\tt NNPDF2.3~PDF}~\cite{Ball:2012cx} for the event generation or {\tt Powheg~Box} interfaced with {\tt Herwig} 7.0.4~\cite{Bahr:2008pv,Bellm:2015jjp} or 7.1.3~\cite{Bellm:2017bvx}. In the relevant plots the following six different options are shown (using the same colour coding as in Ref.~\cite{ATLAS:2023gsl}):
\\[0.5cm]
\begin{itemize*}
  \item[\textcolor{red}{\textbullet}] \textcolor{red}{aMC$@$NLO+Herwig7.1.3} \hspace{4cm}
\item[\textcolor{green}{\textbullet}] \textcolor{green}{Powheg+Herwig7.0.4} \hspace{4cm}
  \item[\textcolor{blue}{\textbullet}] \textcolor{blue}{Powheg+Pythia8} \\[0.2cm]
  \item[\textcolor{magenta}{\textbullet}]\textcolor{magenta}{Powheg+Herwig7.1.3} \hspace{4cm}
  \item[\textcolor{cyan}{\textbullet}] \textcolor{cyan}{aMC$@$NLO+Pythia8} \hspace{4cm}
  \item[\textcolor{dgreen}{\textbullet}] \textcolor{dgreen}{Powheg+Pythia8~(rew.)} 
\end{itemize*} \\[0.1cm]

In the last case, the $p_T$ of the top quark was reweighted using the kinematics of the top quarks after initial- and final-state radiation using next-to-next-to-leading order QCD with next-to-leading EW corrections~\cite{Czakon:2017wor}.\footnote{Note that also the NNLO corrections of Refs.~\cite{Behring:2019iiv,Czakon:2020qbd} are not capable of bringing SM and data in the low $m^{e\mu}$ region into agreement but improve the agreement at high $m^{e\mu}$ where our NP models does not contribute.} The values of $x_i$, are unfortunately not yet available in numerical form and we thus obtained them by digitizing the lower panels of Fig.~10 and 11 using WebPlotDigitizer~v4.6~\cite{Rohatgi2022} with the integrated automatic tool. The results of this extraction for $m^{e\mu}$ and $|\Delta\phi^{e\mu}|$ are shown in Fig.~\ref{fig:mlldata} as dashed lines. 

There are two main sources of correlations. First, there is the correlation between events in the $m^{e\mu}$ bins and in $|\Delta\phi^{e\mu}|$ bins since they are overlapping (i.e.~not mutually exclusive). To estimate this, we simulated 1600k events of $pp\to t\bar t$ in the SM with {\tt MadGraph5aMC@NLO}~\cite{Alwall:2014hca}, {\tt Pythia8.3}~\cite{Sjostrand:2014zea}, and {\tt Delphes}~\cite{deFavereau:2013fsa}. Let $N_i^m$ ($N_\alpha^{\Delta \phi}$) be the number of events in the bin $i^{\rm th}$ ($\alpha^{\rm th}$) of $m^{e\mu}$ ($\Delta\phi^{e\mu}$), disregarding the  $\Delta\phi^{e\mu}$ ($m^{e\mu}$) values. $N_{i \alpha}$ denotes the number of events that are both within $i^{\rm th}$ bin of $m^{e\mu}$ and the $\alpha^{\rm th}$ bin of $\Delta\phi^{e\mu}$. The correlation matrix between $m^{e\mu}$ and $\Delta\phi^{e\mu}$ is then given by
{\begin{equation}
    {\rho _{i\alpha }} = \frac{\delta _i\delta_\alpha{{N_{i\alpha }}}}{{\sqrt {\sum\limits_j {N_j^m} } \sqrt {\sum\limits_\beta  {N_\beta ^\phi } } }}\,.
\end{equation}}
The second source leading to a correlation is due to the normalization of the distributions to the total cross-section. Within a single distribution ($m^{e\mu}$ or $\Delta\phi^{e\mu}$),  one has 
{\begin{equation} 
{\rho _{ij}} = \delta _i\delta_j \left( {\delta _{ij}} - \frac{{2{N_i}{N_j}}}{{N_{}^2}} \right)\,,
\end{equation}}
i.e.~an anti-correlation. With this at hand, we can calculate the $\chi^2$  for the different SM predictions given by ATLAS, see Table~\ref{tab:resmass150}, in the standard way.\footnote{Due to the lack of information, we assumed the systematic errors to have the same correlations as the statistical ones meaning that we applied the correlation matrix to the full error given in the ATLAS tables. Therefore, and due to the digitization accuracy, our values slightly differ from the ones given in the ATLAS paper. However, the impact on the relevant $\Delta \chi^2$ values calculated with a NP hypothesis is expected to be small.}

Next, we can inject an arbitrary NP signal which we normalize to the corresponding total NP cross-section as well as the (normalized) SM cross-section. This means for each (normalized) value of bin $i$ of a given distribution we add a normalized cross-section of
\begin{equation}
    r_i=
    \frac{\sigma_i^{\rm NP}/\sigma^{\rm NP}}
    {\sigma_i^{\rm SM}/\sigma^{\rm SM}}\,,
\end{equation}
with $\sum_i\sigma_i^{\rm NP,SM}=\sigma^{\rm NP,SM}$. We can now calculate the $\chi^2$ including the NP contribution. Treating NP linearly as a small perturbation we have
\begin{equation}
\chi _{{\rm{NP}}}^2 = \sum\limits_{i,j = 1}^{} {{\left( {a{x_i} + {\varepsilon _{{\rm{NP}}}}{r_i} - 1} \right) {\rho^{-1}_{ij}} \left( {a{x_j} + {\varepsilon _{{\rm{NP}}}}{r_j} - 1} \right)}}.
\end{equation}
In order to find the best fit, the $\chi^2$ function is minimized with respect to $\epsilon_{\rm NP}$ and $a$ (i.e. the fitting parameters of the $\chi^2$ distribution), the latter taking into account the possible rescaling of the total cross-section (as done in the ATLAS analysis). Importantly, this means that even if the NP contribution is localized at low $m^{e\mu}$ or $|\Delta \phi^{e\mu}|$ values, higher values are affected since the (predicted) values of $x_i$ are lowered for a non-vanishing NP effect. The values of $r_i$ must now be determined from a NP and SM simulation and the total NP cross-section can be approximated by
\begin{equation}
    \sigma_{\rm NP}\approx\varepsilon _{{\rm{NP}}}\sum_i r_i \sigma^{\rm EXP}_i\,,
\end{equation}
if NP is a small correction to the SM.\footnote{Note that $\epsilon_{\rm NP}$ is only used as a fitting parameter while the results are given in terms of $\sigma_{\rm NP}$, which is equivalent.}

\begin{figure}
    \centering
    \includegraphics[width=1\linewidth]{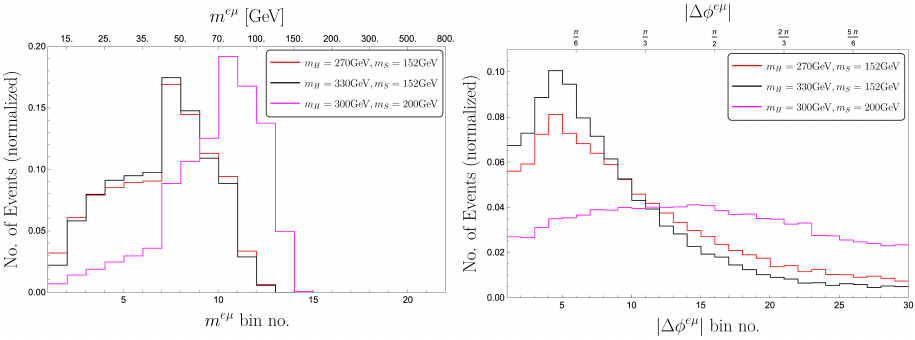}
    \caption{Number of normalized events of the NP process $p p \to H \to S S^\prime \to WW bb $ in the bins of $m^{e \mu}$ (left) and $|\Delta \phi^{e \mu}|$ (right) for three choices of masses: $m_H = 270 \;\text{GeV, } m_S = 152 \; \text{GeV; }$ $m_H = 330 \;\text{GeV, } m_S = 152 \;\text{GeV; }$ $m_H = 300 \;\text{GeV, } m_S = 200 \;\text{GeV; }$ while fixing $m_{S^\prime} = 95 \;\text{GeV}$. The NP distribution appears nearly independent of $m_H$ and is primarily affected by $m_S$.} 
    \label{meu_deltaphi_NPNorm}
\end{figure}

\section{Simplified Model Analysis}
Let us now consider a simplified model in which $H$ is produced via gluon fusion and decays into two lighter scalars $S$ and $S^\prime$ which subsequently decay to $WW$ and $b\bar b$, respectively.\footnote{Note that this is naturally the case if $S$ is a SM-like Higgs and $S^\prime$ is the neutral component of an $SU(2)_L$ triplet with hypercharge~0.} We use the hints for di-photon resonances to fix $m_S=152\,$GeV (for now) and $m_{S^\prime}=95\,$GeV and assume $m_H>m_S+m_{S\prime}$ such that an on-shell decay is possible. For concreteness, we will fix $m_H=270\,$GeV, as suggested in Ref.~\cite{vonBuddenbrock:2017gvy} as a benchmark point, but we checked that varying its mass has a negligible impact on the fit as shown in Fig.~\ref{meu_deltaphi_NPNorm}. 

We simulated in this setup the process $pp\to H\to S S^\prime\to (W W^{(*)}\to \ell^+\nu\ell^-\nu) b\bar b$, with $\ell=e,\mu,\tau$, and the tau lepton subsequently decaying, using {\tt MadGraph5aMC@NLO}~\cite{Alwall:2014hca}, {\tt Pythia8.3}~\cite{Sjostrand:2014zea}, and {\tt Delphes}~\cite{deFavereau:2013fsa}. Note that since our NP contribution is a small perturbation compared to the SM cross-section, the dependence of it on the MC generators used is small. This now determines the shape of the NP contribution to the differential lepton distributions. As outlined in the introduction, we will focus on $m^{e\mu}$ and $|\Delta\phi^{e\mu}|$, including the correlations among them as described in the previous section. The results are shown as solid lines in Fig.~\ref{fig:mlldata}. One can see that for all SM simulations used by ATLAS, the agreement with data is significantly improved. In fact, as given in Table.~\ref{tab:resmass150}, the $\chi^2$ is reduced between $\Delta\chi^2=34$ and $\Delta\chi^2=158$, corresponding to a preference of the NP model over the SM hypothesis by $5.8\sigma$--$13.0\sigma$ while the average of the SM predictions results in $9.6\sigma$. In the appendix, we use the best-fit points for the various SM predictions and their combination to predict the differential distributions. We find in Fig.~\ref{Distributions} that while also here the agreement with data is improved, the effect (reduction of the $\chi^2$) is much less significant than for $m^{e\mu}$ and $|\Delta\phi^{e\mu}|$, justifying our choice of input for the global fit.

\begin{figure}[t]
\begin{center}
    \includegraphics[scale=1]{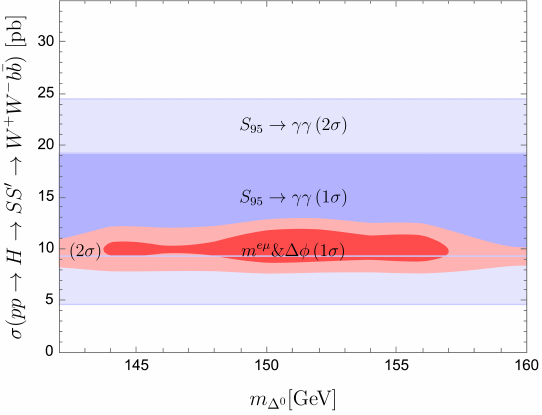}
\end{center}
    \caption{Preferred regions from the $t\bar t$ differential distributions obtained from the average of the MC simulations, i.e.~$9$pb, (red) as a function of $m_S$ and the total cross section $pp\to H\to SS^\prime\to WWbb$ assuming $S^\prime$ to be SM-like and Br$[S\to WW]=100\%$. The blue region is preferred by the $95\,$GeV $\gamma\gamma$ signal strength at 1$\sigma$ and 2$\sigma$.}
    \label{mSvsEpsNP}
\end{figure}

So far, we fixed the masses of $S$ and $S^\prime$ by using the hints for narrow di-photon resonances and took $m_H=270\,$GeV as a benchmark point which avoids constraints from SUSY and non-resonant di-Higgs searches. Let us now discuss the dependence on the masses. First of all, we checked that the effect of changing either the mass of $H$ or of ${S^\prime}$ is very small\footnote{Note that while the effect on the determined NP contribution to the cross section is small since the efficiencies do not depend strongly on $m_H$, the gluon fusion production cross section of $H$ falls with increasing mass. However, this effect can be e.g.~absorbed into its top Yukawa coupling in the case of the UV complete model outlined in Sec.~\ref{sec:conclusions}} which we show in Fig.~\ref{meu_deltaphi_NPNorm}. This means that one can pick any value for $m_H$ so that this does not need to be counted as a degree of freedom in the statistical analysis. Now we can find the best-fit value in the $m_S$-$\epsilon_{\rm NP}$ plane by averaging the $x_i$ values of the six different MC simulations. For this, we generated 500k events each for $m_S$ between 140\,GeV and 160\,GeV in 2\,GeV steps and interpolated. As one can see from the red region in Fig.~\ref{mSvsEpsNP} the preferred $1\sigma$ range, w.r.t.~the best-fit point using two degrees of freedom, for $m_S$ is approximately 144\,GeV -- 157\,GeV, which is perfectly consistent with a mass of $152\,$GeV as suggested by the di-photon excess as around 152\,GeV~\cite{Crivellin:2021ubm}.

Finally, let us consider the consistency of the preferred signal strength for $pp\to H\to SS^\prime$ with the di-photon excess at 95\,GeV. For this, we first fix $\sigma(pp\to H \to S S^\prime)$ using the preferred NP cross-section for $WWbb$ from the average of the MC simulation leading to
\begin{equation}
    \sigma(pp\to H \to S S^\prime)  = \frac{\sigma(pp\to H \to WW b \bar b)}{\text{Br}(S^\prime\to b \bar b) \text{Br}(S \to WW)} \approx \frac{ 9~\text{pb}}{\text{Br}(S^\prime\to b \bar b) \text{Br}(S \to WW)} \,.
\end{equation}
Then assume that $S^\prime$ is SM-like\footnote{The branching ratio to photons of a SM-like Higgs with a mass of $95\,$GeV is $\approx\!1.4\times 10^{-3}$ while the one to $b\bar b$ is $\approx\!0.86$~\cite{LHCHiggsCrossSectionWorkingGroup:2016ypw} and the production cross section at $13\,$TeV via gluon fusion is $\approx\!68$pb~\cite{LHCHiggsCrossSectionWorkingGroup:2016ypw}.} and that $S$ decays to $\approx$100\% to $WW$. We can now calculate the di-photon signal strength of the 95\,GeV Higgs in our model
\begin{equation}
    \mu_{\gamma\gamma}^{\text{NP}} = \frac{\sigma(pp\to H \to S S_{\text{SM}}^\prime)\text{Br}(S_{\text{SM}}^\prime\to\gamma\gamma)}{\sigma(pp\to h_{95})\text{Br}(h_{95}\to\gamma\gamma)} \approx 0.15 \,
\end{equation}
where $h_{95}$ represents a hypothetical Higgs boson with a mass of $95$~GeV\footnote{Notice that this means Br($S_{\text{SM}}^\prime\to\gamma\gamma$) = Br($h_{95}\to\gamma\gamma$) }. Comparing this to the experimental value of $\mu_{\gamma\gamma}=0.24^{+0.09}_{-0.08}$~\cite{Biekotter:2023oen}, one can see in Fig.~\ref{mSvsEpsNP} that the predicted signal strength using the $t\bar t$ differential distributions (red) is perfectly consistent with the region preferred by the di-photon excess at 95\,GeV (blue) in our simplified model. For this prediction, we assumed Br$(S \to WW)\approx 100\%$. Note that while in a UV complete model, one cannot expect that $S$ decays only to $WW$, if the mixing angle $\alpha$ of the neutral triplet component with the SM Higgs is small this decay is dominant.\footnote{{To be more specific if $-0.05<\alpha<0.02$,  Br$(S\to WW)>90\%$ for a triplet vacuum expectation value of 3.4\,GeV~\cite{Ashanujjaman:2024lnr}.}}

\section{Conclusions}
\label{sec:conclusions}

ATLAS found that the measured differential lepton distributions in its $t\bar t$ analysis significantly deviate from the SM predictions obtained for different combinations of simulators: ``No model (SM simulation) can describe all measured distributions within their uncertainties.'' Taking into account the performance of the ATLAS detector w.r.t.~leptons and the level of sophistication of the SM simulations, this suggests that the measurement is contaminated by NP contributions.

In this article, we propose that the process $pp\to H\to SS^\prime \to WW b\bar b$ constitutes a NP background to the measurements of $t\bar t$ differential lepton distributions. Motivated by the hints for di-photon resonances $\approx\!95\,$GeV and $\approx\!152\,$GeV we considered the benchmark point with $S^\prime$ and $S$ having the corresponding masses and decaying to $b\bar b$ and $WW$, respectively (with $m_H=270\,$GeV). We find that this NP hypothesis is preferred over the SM one by $5.8\sigma$ to $13\sigma$, depending on which of the six SM simulations employed by ATLAS is considered. In fact, all differential distributions, in particular $m^{e\mu}$ and $|\Delta \phi^{e\mu}|$, are much better described once NP is included and averaging the SM predictions a significance of $9.6\sigma$ is found.

Varying $m_S$, to which the distributions are most sensitive among the new scalar masses, we showed that the preferred range is compatible with $m_S\approx 152\,$GeV, as motivated by the $\gamma\gamma$ and $WW$ excesses. Since the latter analysis employs a jet veto, this further disfavours the possibility that higher-order QCD corrections are the origin of the tension in the $t\bar t$ differential distributions. Furthermore, averaging the six different SM predictions $\sigma(pp\to H\to SS^\prime \to WW b\bar b)\approx 9$pb is preferred by $t\bar t $ data. Assuming that $S^\prime$ is SM-like, this results in a di-photon signal strength in agreement with the $\gamma\gamma$ excess. This provides further support for the emerging excess at $\approx \,$95\,GeV which naturally has a large $b \bar b$ branching ratio and opens a window of opportunity to further explore this boson in associated production. 

\begin{table}[t]
\begin{center}
\begin{tblr}{colspec={Q[c,1.8cm]|Q[c,0.5cm]Q[c,1.5cm]Q[c,1.5cm]Q[c,1.5cm]}}
{}& $S^\prime$ &${{H_1}}$&${{H_2}}$&$\Delta$ \\
\hline
$SU( 2 )_L$&1&2&2&3\\
$U( 1 )_Y$& 0&{1/2}&{1/2}&0\\
$Z_2^1$& +&-&+&-\\
$Z_2^2$& -&+&-&+\\
components &$S^\prime$&$H,A,H^\pm$&$h,{G^0},{G^\pm }$&$S,\Delta^ \pm$
\end{tblr}
\caption{Fields and their components in the limit of vanishing mixing angles. $h$, $S$ and $S^\prime$ correspond to SM Higgs, triplet Higgs with mass of $\approx150$\,GeV, and singlet Higgs with mass $95$\,GeV, respectively. $G^0$ and $G^\pm$ are the Goldstone bosons while $CP$-even $H$ and $CP$-odd $A$ originate from $H_1$ and have masses around 300\,GeV and 400\,GeV, respectively. Finally, also $H^\pm$ could have a mass around 400\,GeV.}
\label{fields}
\end{center}
\end{table}

Finally, we notice that for $S$ a dominant decay to $W$ bosons is needed while the $ZZ$ rate should be low to respect the limits from inclusive $4\ell$ searches. This suggests that $S$ is the neutral component of the $SU(2)_L$ triplet with hypercharge 0 ($\Delta$), as also motivated by the $W$-mass average. 
In fact, a possible model giving the phenomenology described above is a two-Higgs-doublets model (with the Higgs doublets $H_1$ and $H_2$) supplemented by a triplet with hypercharge 0 ($\Delta$) and a singlet ($S^\prime$), as shown in Table~\ref{fields}. For completeness, let us write the lagrangian sector involving the scalars. As in the 2HDMS, one can employ two $Z_2$ symmetries, one ($Z_2^1$) under which $S^\prime$ and $H_2$ are odd, and one ($Z_2^2$) under which $H_1$ and $\Delta$ are odd, while all other fields are even apart from the right-handed fermions. These $Z_2$ symmetries allow only for Yukawa couplings of $H_2$ and constrain the scalar potential to be

\begin{equation}
\begin{aligned}
    \mathcal{L}_{\rm{scalar}} =& - \mu_1 |H_1|^2 -\mu_2 |H_2|^2 -(\mu_3 H_1^\dag H_2 + {\rm{h.c.}})- \mu_4 {\rm{Tr}}[\Delta^2] - \mu_5 S^{\prime2} + \lambda_{1} |H_1|^4 + \lambda_{2} |H_2|^4 
    \\ & + \lambda_{3} \rm{Tr}[\Delta^4] + \lambda_{4} S^{\prime4} +\lambda_{5} |H_1|^2 |H_2|^2 + \lambda_{6} H_1^\dag H_2 H_2^\dag H_1 + \lambda_{7} ((H_1^\dag H_2)^2 + \rm{h.c.}) + \lambda_{8} |H_1|^2 S^{\prime2}
    \\ &+  \lambda_{9} |H_2|^2 S^{\prime2}+ (\lambda_{10} |H_1|^2+ \lambda_{11} |H_2|^2) \rm{Tr}[\Delta^2] + \lambda_{13}\rm{Tr}[\Delta^2]S^{\prime2} + (\kappa H_1^\dag \Delta H_2 S^\prime + \rm{h.c.})\,. 
    \end{aligned}
\end{equation}

The last term, namely the one proportional to $\kappa$, can generate the desired associated production of $S$ and $S^\prime$ if {$H$ (contained in $H_1$)} is produced via gluon fusion, e.g.~via a top quark loop. We therefore assume that the $Z_2$ symmetries are broken in the top sector by the term
\begin{equation}
    {-\mathcal{L}_Y^{\cancel{Z}_2} = \mu_t \sqrt{2}\dfrac{m_t}{v}  \bar Q_3 \widetilde H_1 u_3\,.}   
\end{equation} 

Furthermore, since $v_\Delta,v_1 \ll v_2$ (where $v_\Delta, v_1, v_2$ are the vacuum expectation values of $\Delta,\, H_1$ and $H_2$, respectively), this naturally leads to a small mixing with the SM-like Higgs (contained in $H_2$) such that the constraints from signal strength measurements are satisfied.\footnote{This model thus contains in addition to the SM Higgs three neutral CP-even Higgses, one neutral CP-odd Higgs and a charged scalar. Note that the interactions of these fields with gauge bosons are determined by their representations under the SM gauge group. In Ref.~\cite{Coloretti:2023yyq}, we recently investigated this model in more detail, using the results of this article from the differential $t\bar t$ distributions as crucial input.}

\begin{acknowledgments}
The work of A.C., S.B.~and G.C.~is supported by a professorship grant from the Swiss National Science Foundation (No.\ PP00P2\_211002). B.M.~gratefully acknowledges the South African Department of Science and Innovation through the SA-CERN program, the National Research Foundation, and the Research Office of the University of the Witwatersrand for various forms of support.
\end{acknowledgments}

\newpage
\appendix
\section{Additional Differential Distributions}
In this Appendix, we present the SM and NP predictions for the four observables $\eta$, $E^e+E^\mu$, $p_T^e+p_T^\mu$ and $p_T^{e \mu}$.
\begin{figure*}[h!]
\begin{centering}
    \includegraphics[width=0.76\linewidth]{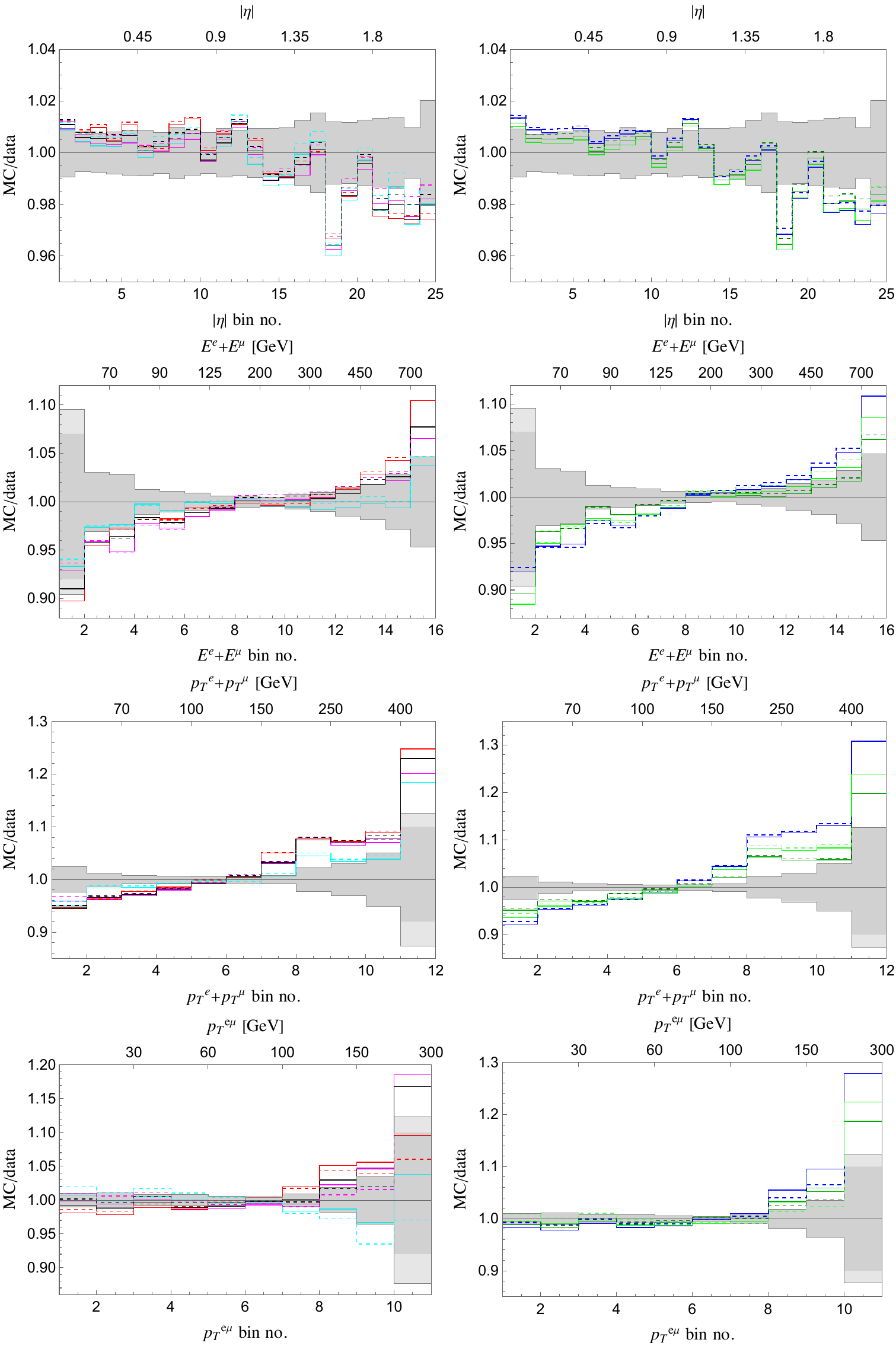}
                   \raisebox{0.9\height}{\includegraphics[width=0.21\linewidth]{Plot_Legend.pdf}}
                    \end{centering}
    \caption{SM values (dashed) and NP predictions (solid), using the best fit from $m^{e\mu}$ and $|\Delta\phi^{e\mu}|$, for six other differential distributions given in the ATLAS paper and their average (black). One can see that while NP improves the agreement of theory with data, the effect is not as significant as in $m^{e\mu}$ and $|\Delta\phi^{e\mu}|$, justifying our input choice. Only $y^{e\mu}$ and $p_T^{e\mu}$ are not shown since here both the tensions within the SM as well as the NP effect are very small.}
    \label{Distributions}
\end{figure*}
\newpage

\bibliographystyle{utphys}
\bibliography{bib}

\providecommand{\href}[2]{#2}\begingroup\raggedright\begin{thebibliography}{10}

\bibitem{ParticleDataGroup:2020ssz}
{\bfseries Particle Data Group} Collaboration, P.~A. Zyla {\em et~al.},
  ``{Review of Particle Physics},''
  \href{https://dx.doi.org/10.1093/ptep/ptaa104}{{\em PTEP} {\bfseries 2020}
  no.~8, (2020) 083C01}.

\bibitem{Higgs:1964ia}
P.~W. Higgs, ``{Broken symmetries, massless particles and gauge fields},''
  \href{https://dx.doi.org/10.1016/0031-9163(64)91136-9}{{\em Phys. Lett.}
  {\bfseries 12} (1964) 132--133}.

\bibitem{Englert:1964et}
F.~Englert and R.~Brout, ``{Broken Symmetry and the Mass of Gauge Vector
  Mesons},'' \href{https://dx.doi.org/10.1103/PhysRevLett.13.321}{{\em Phys.
  Rev. Lett.} {\bfseries 13} (1964) 321--323}.

\bibitem{Higgs:1964pj}
P.~W. Higgs, ``{Broken Symmetries and the Masses of Gauge Bosons},''
  \href{https://dx.doi.org/10.1103/PhysRevLett.13.508}{{\em Phys. Rev. Lett.}
  {\bfseries 13} (1964) 508--509}.

\bibitem{Guralnik:1964eu}
G.~S. Guralnik, C.~R. Hagen, and T.~W.~B. Kibble, ``{Global Conservation Laws
  and Massless Particles},''
  \href{https://dx.doi.org/10.1103/PhysRevLett.13.585}{{\em Phys. Rev. Lett.}
  {\bfseries 13} (1964) 585--587}.

\bibitem{Aad:2012tfa}
{\bfseries ATLAS} Collaboration, G.~Aad {\em et~al.}, ``{Observation of a new
  particle in the search for the Standard Model Higgs boson with the ATLAS
  detector at the LHC},''
  \href{https://dx.doi.org/10.1016/j.physletb.2012.08.020}{{\em Phys. Lett. B}
  {\bfseries 716} (2012) 1--29},
  \href{https://arxiv.org/abs/1207.7214}{{\ttfamily arXiv:1207.7214 [hep-ex]}}.

\bibitem{Chatrchyan:2012ufa}
{\bfseries CMS} Collaboration, S.~Chatrchyan {\em et~al.}, ``{Observation of a
  New Boson at a Mass of 125 GeV with the CMS Experiment at the LHC},''
  \href{https://dx.doi.org/10.1016/j.physletb.2012.08.021}{{\em Phys. Lett. B}
  {\bfseries 716} (2012) 30--61},
  \href{https://arxiv.org/abs/1207.7235}{{\ttfamily arXiv:1207.7235 [hep-ex]}}.

\bibitem{Langford:2021osp}
{\bfseries ATLAS, CMS} Collaboration, J.~M. Langford, ``{Combination of Higgs
  measurements from ATLAS and CMS : couplings and $\mathcal{k}$- framework},''
  \href{https://dx.doi.org/10.22323/1.382.0136}{{\em PoS} {\bfseries LHCP2020}
  (2021) 136}.

\bibitem{ATLAS:2021vrm}
{\bfseries ATLAS} Collaboration, ``{Combined measurements of Higgs boson
  production and decay using up to $139$ fb$^{-1}$ of proton-proton collision
  data at $\sqrt{s}= 13$ TeV collected with the ATLAS experiment},''.

\bibitem{ALEPH:2013dgf}
{\bfseries ALEPH, DELPHI, L3, OPAL, LEP Electroweak} Collaboration, S.~Schael
  {\em et~al.}, ``{Electroweak Measurements in Electron-Positron Collisions at
  W-Boson-Pair Energies at LEP},''
  \href{https://dx.doi.org/10.1016/j.physrep.2013.07.004}{{\em Phys. Rept.}
  {\bfseries 532} (2013) 119--244},
  \href{https://arxiv.org/abs/1302.3415}{{\ttfamily arXiv:1302.3415 [hep-ex]}}.

\bibitem{LHCb:2021bjt}
{\bfseries LHCb} Collaboration, R.~Aaij {\em et~al.}, ``{Measurement of the W
  boson mass},'' \href{https://dx.doi.org/10.1007/JHEP01(2022)036}{{\em JHEP}
  {\bfseries 01} (2022) 036},
  \href{https://arxiv.org/abs/2109.01113}{{\ttfamily arXiv:2109.01113
  [hep-ex]}}.

\bibitem{CDF:2022hxs}
{\bfseries CDF} Collaboration, T.~Aaltonen {\em et~al.}, ``{High-precision
  measurement of the $W$ boson mass with the CDF II detector},''
  \href{https://dx.doi.org/10.1126/science.abk1781}{{\em Science} {\bfseries
  376} no.~6589, (2022) 170--176}.

\bibitem{ATLAS:2023fsi}
{\bfseries ATLAS} Collaboration, ``{Improved W boson Mass Measurement using 7
  TeV Proton-Proton Collisions with the ATLAS Detector},''.

\bibitem{LHC-TeVMWWorkingGroup:2023zkn}
{\bfseries LHC-TeV~MW~Working~Group} Collaboration, S.~Amoroso {\em et~al.},
  ``{Compatibility and combination of world W-boson mass measurements},''
  \href{https://dx.doi.org/10.1140/epjc/s10052-024-12532-z}{{\em Eur. Phys. J.
  C} {\bfseries 84} no.~5, (2024) 451},
  \href{https://arxiv.org/abs/2308.09417}{{\ttfamily arXiv:2308.09417
  [hep-ex]}}.

\bibitem{Strumia:2022qkt}
A.~Strumia, ``{Interpreting electroweak precision data including the W-mass CDF
  anomaly},'' \href{https://dx.doi.org/10.1007/JHEP08(2022)248}{{\em JHEP}
  {\bfseries 08} (2022) 248},
  \href{https://arxiv.org/abs/2204.04191}{{\ttfamily arXiv:2204.04191
  [hep-ph]}}.

\bibitem{LEPWorkingGroupforHiggsbosonsearches:2003ing}
{\bfseries LEP Working Group for Higgs boson searches, ALEPH, DELPHI, L3, OPAL}
  Collaboration, R.~Barate {\em et~al.}, ``{Search for the standard model Higgs
  boson at LEP},'' \href{https://dx.doi.org/10.1016/S0370-2693(03)00614-2}{{\em
  Phys. Lett. B} {\bfseries 565} (2003) 61--75},
  \href{https://arxiv.org/abs/hep-ex/0306033}{{\ttfamily
  arXiv:hep-ex/0306033}}.

\bibitem{CMS:2018cyk}
{\bfseries CMS} Collaboration, A.~M. Sirunyan {\em et~al.}, ``{Search for a
  standard model-like Higgs boson in the mass range between 70 and 110 GeV in
  the diphoton final state in proton-proton collisions at $\sqrt{s}=$ 8 and 13
  TeV},'' \href{https://dx.doi.org/10.1016/j.physletb.2019.03.064}{{\em Phys.
  Lett. B} {\bfseries 793} (2019) 320--347},
  \href{https://arxiv.org/abs/1811.08459}{{\ttfamily arXiv:1811.08459
  [hep-ex]}}.

\bibitem{CMS:2022rbd}
{\bfseries CMS} Collaboration, A.~M. Sirunyan {\em et~al.}, ``{Searches for
  additional Higgs bosons and vector leptoquarks in $\tau\tau$ final states in
  proton-proton collisions at $\sqrt{s}=13~\mathrm{TeV}$},''.
  \url{http://cds.cern.ch/record/2803739}.

\bibitem{CMS:2022tgk}
{\bfseries CMS} Collaboration, A.~M. Sirunyan {\em et~al.}, ``{Search for a new
  resonance decaying to two scalars in the final state with two bottom quarks
  and two photons in proton-proton collisions at
  $\sqrt{s}=13\,\mathrm{TeV}$},''.

\bibitem{ATLAS:2023jzc}
{\bfseries ATLAS} Collaboration, G.~Aad {\em et~al.}, ``{Search for diphoton
  resonances in the 66 to 110 GeV mass range using 140 fb$^{-1}$ of 13 TeV $pp$
  collisions collected with the ATLAS detector},''.

\bibitem{ATLAS:2021jbf}
{\bfseries ATLAS} Collaboration, G.~Aad {\em et~al.}, ``{Search for dark matter
  in events with missing transverse momentum and a Higgs boson decaying into
  two photons in pp collisions at $ \sqrt{s} $ = 13 TeV with the ATLAS
  detector},'' \href{https://dx.doi.org/10.1007/JHEP10(2021)013}{{\em JHEP}
  {\bfseries 10} (2021) 013},
  \href{https://arxiv.org/abs/2104.13240}{{\ttfamily arXiv:2104.13240
  [hep-ex]}}.

\bibitem{Crivellin:2021ubm}
A.~Crivellin, Y.~Fang, O.~Fischer, A.~Kumar, M.~Kumar, E.~Malwa, B.~Mellado,
  N.~Rapheeha, X.~Ruan, and Q.~Sha, ``{Accumulating Evidence for the Associate
  Production of a Neutral Scalar with Mass around 151 GeV},'' {\em accepted for
  publication in Phys. Rev. D} (9, 2021) ,
  \href{https://arxiv.org/abs/2109.02650}{{\ttfamily arXiv:2109.02650
  [hep-ph]}}.

\bibitem{Bhattacharya:2023lmu}
S.~Bhattacharya, G.~Coloretti, A.~Crivellin, S.-E. Dahbi, Y.~Fang, M.~Kumar,
  and B.~Mellado, ``{Growing Excesses of New Scalars at the Electroweak
  Scale},'' \href{https://arxiv.org/abs/2306.17209}{{\ttfamily arXiv:2306.17209
  [hep-ph]}}.

\bibitem{Buddenbrock:2019tua}
S.~Buddenbrock, A.~S. Cornell, Y.~Fang, A.~Fadol~Mohammed, M.~Kumar,
  B.~Mellado, and K.~G. Tomiwa, ``{The emergence of multi-lepton anomalies at
  the LHC and their compatibility with new physics at the EW scale},''
  \href{https://dx.doi.org/10.1007/JHEP10(2019)157}{{\em JHEP} {\bfseries 10}
  (2019) 157}, \href{https://arxiv.org/abs/1901.05300}{{\ttfamily
  arXiv:1901.05300 [hep-ph]}}.

\bibitem{vonBuddenbrock:2020ter}
S.~von Buddenbrock, R.~Ruiz, and B.~Mellado, ``{Anatomy of inclusive $t\bar t
  W$ production at hadron colliders},''
  \href{https://dx.doi.org/10.1016/j.physletb.2020.135964}{{\em Phys. Lett. B}
  {\bfseries 811} (2020) 135964},
  \href{https://arxiv.org/abs/2009.00032}{{\ttfamily arXiv:2009.00032
  [hep-ph]}}.

\bibitem{Hernandez:2019geu}
Y.~Hernandez, M.~Kumar, A.~S. Cornell, S.-E. Dahbi, Y.~Fang, B.~Lieberman,
  B.~Mellado, K.~Monnakgotla, X.~Ruan, and S.~Xin, ``{The anomalous production
  of multi-lepton and its impact on the measurement of $Wh$ production at the
  LHC},'' \href{https://dx.doi.org/10.1140/epjc/s10052-021-09137-1}{{\em Eur.
  Phys. J. C} {\bfseries 81} no.~4, (2021) 365},
  \href{https://arxiv.org/abs/1912.00699}{{\ttfamily arXiv:1912.00699
  [hep-ph]}}.

\bibitem{Fischer:2021sqw}
O.~Fischer {\em et~al.}, ``{Unveiling hidden physics at the LHC},''
  \href{https://dx.doi.org/10.1140/epjc/s10052-022-10541-4}{{\em Eur. Phys. J.
  C} {\bfseries 82} no.~8, (2022) 665},
  \href{https://arxiv.org/abs/2109.06065}{{\ttfamily arXiv:2109.06065
  [hep-ph]}}.

\bibitem{vonBuddenbrock:2017gvy}
S.~von Buddenbrock, A.~S. Cornell, A.~Fadol, M.~Kumar, B.~Mellado, and X.~Ruan,
  ``{Multi-lepton signatures of additional scalar bosons beyond the Standard
  Model at the LHC},'' \href{https://dx.doi.org/10.1088/1361-6471/aae3d6}{{\em
  J. Phys. G} {\bfseries 45} no.~11, (2018) 115003},
  \href{https://arxiv.org/abs/1711.07874}{{\ttfamily arXiv:1711.07874
  [hep-ph]}}.

\bibitem{Coloretti:2023wng}
G.~Coloretti, A.~Crivellin, S.~Bhattacharya, and B.~Mellado, ``{Searching for
  Low-Mass Resonances Decaying into $W$ Bosons},''
  \href{https://arxiv.org/abs/2302.07276}{{\ttfamily arXiv:2302.07276
  [hep-ph]}}.

\bibitem{ATLAS:2023gsl}
{\bfseries ATLAS} Collaboration, G.~Aad {\em et~al.}, ``{Inclusive and
  differential cross-sections for dilepton $ t\overline{t} $ production
  measured in $ \sqrt{s} $ = 13 TeV pp collisions with the ATLAS detector},''
  \href{https://dx.doi.org/10.1007/JHEP07(2023)141}{{\em JHEP} {\bfseries 07}
  (2023) 141}, \href{https://arxiv.org/abs/2303.15340}{{\ttfamily
  arXiv:2303.15340 [hep-ex]}}.

\bibitem{Beneke:2011mq}
M.~Beneke, P.~Falgari, S.~Klein, and C.~Schwinn, ``{Hadronic top-quark pair
  production with NNLL threshold resummation},''
  \href{https://dx.doi.org/10.1016/j.nuclphysb.2011.10.021}{{\em Nucl. Phys. B}
  {\bfseries 855} (2012) 695--741},
  \href{https://arxiv.org/abs/1109.1536}{{\ttfamily arXiv:1109.1536 [hep-ph]}}.

\bibitem{Barnreuther:2012wtj}
P.~B\"arnreuther, M.~Czakon, and A.~Mitov, ``{Percent Level Precision Physics
  at the Tevatron: First Genuine NNLO QCD Corrections to $q \bar{q} \to t
  \bar{t} + X$},''
  \href{https://dx.doi.org/10.1103/PhysRevLett.109.132001}{{\em Phys. Rev.
  Lett.} {\bfseries 109} (2012) 132001},
  \href{https://arxiv.org/abs/1204.5201}{{\ttfamily arXiv:1204.5201 [hep-ph]}}.

\bibitem{Czakon:2012zr}
M.~Czakon and A.~Mitov, ``{NNLO corrections to top-pair production at hadron
  colliders: the all-fermionic scattering channels},''
  \href{https://dx.doi.org/10.1007/JHEP12(2012)054}{{\em JHEP} {\bfseries 12}
  (2012) 054}, \href{https://arxiv.org/abs/1207.0236}{{\ttfamily
  arXiv:1207.0236 [hep-ph]}}.

\bibitem{Kidonakis:2023jpj}
N.~Kidonakis and C.~Foster, ``{Soft-gluon corrections in $t{\bar t}W$
  production},'' \href{https://arxiv.org/abs/2312.00861}{{\ttfamily
  arXiv:2312.00861 [hep-ph]}}.

\bibitem{ATLAS:2017dhr}
{\bfseries ATLAS} Collaboration, M.~Aaboud {\em et~al.}, ``{Measurement of
  lepton differential distributions and the top quark mass in $t\bar{t}$
  production in $pp$ collisions at $\sqrt{s}=8$ TeV with the ATLAS detector},''
  \href{https://dx.doi.org/10.1140/epjc/s10052-017-5349-9}{{\em Eur. Phys. J.
  C} {\bfseries 77} no.~11, (2017) 804},
  \href{https://arxiv.org/abs/1709.09407}{{\ttfamily arXiv:1709.09407
  [hep-ex]}}.

\bibitem{ATLAS:2019hau}
{\bfseries ATLAS} Collaboration, G.~Aad {\em et~al.}, ``{Measurement of the
  $t\bar{t}$ production cross-section and lepton differential distributions in
  $e\mu $ dilepton events from $pp$ collisions at $\sqrt{s}=13\,\text {TeV}$
  with the ATLAS detector},''
  \href{https://dx.doi.org/10.1140/epjc/s10052-020-7907-9}{{\em Eur. Phys. J.
  C} {\bfseries 80} no.~6, (2020) 528},
  \href{https://arxiv.org/abs/1910.08819}{{\ttfamily arXiv:1910.08819
  [hep-ex]}}.

\bibitem{CMS:2018adi}
{\bfseries CMS} Collaboration, A.~M. Sirunyan {\em et~al.}, ``{Measurements of
  $\mathrm{t\overline{t}}$ differential cross sections in proton-proton
  collisions at $\sqrt{s}=$ 13 TeV using events containing two leptons},''
  \href{https://dx.doi.org/10.1007/JHEP02(2019)149}{{\em JHEP} {\bfseries 02}
  (2019) 149}, \href{https://arxiv.org/abs/1811.06625}{{\ttfamily
  arXiv:1811.06625 [hep-ex]}}.

\bibitem{CMS:2020djy}
{\bfseries CMS} Collaboration, A.~M. Sirunyan {\em et~al.}, ``{Measurement of
  the top quark Yukawa coupling from $\mathrm{t\bar{t}}$ kinematic
  distributions in the dilepton final state in proton-proton collisions at
  $\sqrt{s}=$ 13 TeV},''
  \href{https://dx.doi.org/10.1103/PhysRevD.102.092013}{{\em Phys. Rev. D}
  {\bfseries 102} no.~9, (2020) 092013},
  \href{https://arxiv.org/abs/2009.07123}{{\ttfamily arXiv:2009.07123
  [hep-ex]}}.

\bibitem{CMS:2018amb}
{\bfseries CMS} Collaboration, A.~M. Sirunyan {\em et~al.}, ``{Measurement of
  the production cross section for single top quarks in association with W
  bosons in proton-proton collisions at $ \sqrt{s}=13 $ TeV},''
  \href{https://dx.doi.org/10.1007/JHEP10(2018)117}{{\em JHEP} {\bfseries 10}
  (2018) 117}, \href{https://arxiv.org/abs/1805.07399}{{\ttfamily
  arXiv:1805.07399 [hep-ex]}}.

\bibitem{ATLAS:2010arf}
{\bfseries ATLAS} Collaboration, G.~Aad {\em et~al.}, ``{The ATLAS Simulation
  Infrastructure},''
  \href{https://dx.doi.org/10.1140/epjc/s10052-010-1429-9}{{\em Eur. Phys. J.
  C} {\bfseries 70} (2010) 823--874},
  \href{https://arxiv.org/abs/1005.4568}{{\ttfamily arXiv:1005.4568
  [physics.ins-det]}}.

\bibitem{CMS:2019csb}
{\bfseries CMS} Collaboration, A.~M. Sirunyan {\em et~al.}, ``{Extraction and
  validation of a new set of CMS PYTHIA8 tunes from underlying-event
  measurements},''
  \href{https://dx.doi.org/10.1140/epjc/s10052-019-7499-4}{{\em Eur. Phys. J.
  C} {\bfseries 80} no.~1, (2020) 4},
  \href{https://arxiv.org/abs/1903.12179}{{\ttfamily arXiv:1903.12179
  [hep-ex]}}.

\bibitem{ATLAS:2019qmc}
{\bfseries ATLAS} Collaboration, G.~Aad {\em et~al.}, ``{Electron and photon
  performance measurements with the ATLAS detector using the
  2015\textendash{}2017 LHC proton-proton collision data},''
  \href{https://dx.doi.org/10.1088/1748-0221/14/12/P12006}{{\em JINST}
  {\bfseries 14} no.~12, (2019) P12006},
  \href{https://arxiv.org/abs/1908.00005}{{\ttfamily arXiv:1908.00005
  [hep-ex]}}.

\bibitem{CMS:2015xaf}
{\bfseries CMS} Collaboration, V.~Khachatryan {\em et~al.}, ``{Performance of
  Electron Reconstruction and Selection with the CMS Detector in Proton-Proton
  Collisions at \ensuremath{\sqrt{}}s = 8 TeV},''
  \href{https://dx.doi.org/10.1088/1748-0221/10/06/P06005}{{\em JINST}
  {\bfseries 10} no.~06, (2015) P06005},
  \href{https://arxiv.org/abs/1502.02701}{{\ttfamily arXiv:1502.02701
  [physics.ins-det]}}.

\bibitem{CMS:2012nsv}
{\bfseries CMS} Collaboration, S.~Chatrchyan {\em et~al.}, ``{Performance of
  CMS Muon Reconstruction in $pp$ Collision Events at $\sqrt{s}=7$ TeV},''
  \href{https://dx.doi.org/10.1088/1748-0221/7/10/P10002}{{\em JINST}
  {\bfseries 7} (2012) P10002},
  \href{https://arxiv.org/abs/1206.4071}{{\ttfamily arXiv:1206.4071
  [physics.ins-det]}}.

\bibitem{CMS:2022uhn}
{\bfseries CMS} Collaboration, A.~Tumasyan {\em et~al.}, ``{Measurements of the
  Higgs boson production cross section and couplings in the W boson pair decay
  channel in proton-proton collisions at $\sqrt{s}=13\,\text
  {Te\hspace{-.08em}V} $},''
  \href{https://dx.doi.org/10.1140/epjc/s10052-023-11632-6}{{\em Eur. Phys. J.
  C} {\bfseries 83} no.~7, (2023) 667},
  \href{https://arxiv.org/abs/2206.09466}{{\ttfamily arXiv:2206.09466
  [hep-ex]}}.

\bibitem{ATLAS:2022ooq}
{\bfseries ATLAS} Collaboration, ``{Measurements of Higgs boson production by
  gluon$-$gluon fusion and vector-boson fusion using $H\rightarrow W W^*
  \rightarrow e\nu \mu\nu$ decays in $pp$ collisions at $\sqrt{s}=13$ TeV with
  the ATLAS detector},'' \href{https://arxiv.org/abs/2207.00338}{{\ttfamily
  arXiv:2207.00338 [hep-ex]}}.

\bibitem{Jezo:2023rht}
T.~Je\v{z}o, J.~M. Lindert, and S.~Pozzorini, ``{Resonance-aware NLOPS matching
  for off-shell $t\bar t+tW$ production with semileptonic decays},''
  \href{https://arxiv.org/abs/2307.15653}{{\ttfamily arXiv:2307.15653
  [hep-ph]}}.

\bibitem{Frixione:2007vw}
S.~Frixione, P.~Nason, and C.~Oleari, ``{Matching NLO QCD computations with
  Parton Shower simulations: the POWHEG method},''
  \href{https://dx.doi.org/10.1088/1126-6708/2007/11/070}{{\em JHEP} {\bfseries
  11} (2007) 070}, \href{https://arxiv.org/abs/0709.2092}{{\ttfamily
  arXiv:0709.2092 [hep-ph]}}.

\bibitem{Alioli:2010xd}
S.~Alioli, P.~Nason, C.~Oleari, and E.~Re, ``{A general framework for
  implementing NLO calculations in shower Monte Carlo programs: the POWHEG
  BOX},'' \href{https://dx.doi.org/10.1007/JHEP06(2010)043}{{\em JHEP}
  {\bfseries 06} (2010) 043}, \href{https://arxiv.org/abs/1002.2581}{{\ttfamily
  arXiv:1002.2581 [hep-ph]}}.

\bibitem{Frixione:2007nw}
S.~Frixione, P.~Nason, and G.~Ridolfi, ``{A Positive-weight
  next-to-leading-order Monte Carlo for heavy flavour hadroproduction},''
  \href{https://dx.doi.org/10.1088/1126-6708/2007/09/126}{{\em JHEP} {\bfseries
  09} (2007) 126}, \href{https://arxiv.org/abs/0707.3088}{{\ttfamily
  arXiv:0707.3088 [hep-ph]}}.

\bibitem{NNPDF:2014otw}
{\bfseries NNPDF} Collaboration, R.~D. Ball {\em et~al.}, ``{Parton
  distributions for the LHC Run II},''
  \href{https://dx.doi.org/10.1007/JHEP04(2015)040}{{\em JHEP} {\bfseries 04}
  (2015) 040}, \href{https://arxiv.org/abs/1410.8849}{{\ttfamily
  arXiv:1410.8849 [hep-ph]}}.

\bibitem{Sjostrand:2006za}
T.~Sjostrand, S.~Mrenna, and P.~Z. Skands, ``{PYTHIA 6.4 Physics and Manual},''
  \href{https://dx.doi.org/10.1088/1126-6708/2006/05/026}{{\em JHEP} {\bfseries
  05} (2006) 026}, \href{https://arxiv.org/abs/hep-ph/0603175}{{\ttfamily
  arXiv:hep-ph/0603175}}.

\bibitem{Sjostrand:2014zea}
T.~Sj\"ostrand, S.~Ask, J.~R. Christiansen, R.~Corke, N.~Desai, P.~Ilten,
  S.~Mrenna, S.~Prestel, C.~O. Rasmussen, and P.~Z. Skands, ``{An introduction
  to PYTHIA 8.2}'' \href{https://dx.doi.org/10.1016/j.cpc.2015.01.024}{{\em
  Comput. Phys. Commun.} {\bfseries 191} (2015) 159--177},
  \href{https://arxiv.org/abs/1410.3012}{{\ttfamily arXiv:1410.3012 [hep-ph]}}.

\bibitem{Alwall:2014hca}
J.~Alwall, R.~Frederix, S.~Frixione, V.~Hirschi, F.~Maltoni, O.~Mattelaer,
  H.~S. Shao, T.~Stelzer, P.~Torrielli, and M.~Zaro, ``{The automated
  computation of tree-level and next-to-leading order differential cross
  sections, and their matching to parton shower simulations},''
  \href{https://dx.doi.org/10.1007/JHEP07(2014)079}{{\em JHEP} {\bfseries 07}
  (2014) 079}, \href{https://arxiv.org/abs/1405.0301}{{\ttfamily
  arXiv:1405.0301 [hep-ph]}}.

\bibitem{Ball:2012cx}
R.~D. Ball {\em et~al.}, ``{Parton distributions with LHC data},''
  \href{https://dx.doi.org/10.1016/j.nuclphysb.2012.10.003}{{\em Nucl. Phys. B}
  {\bfseries 867} (2013) 244--289},
  \href{https://arxiv.org/abs/1207.1303}{{\ttfamily arXiv:1207.1303 [hep-ph]}}.

\bibitem{Bahr:2008pv}
M.~Bahr {\em et~al.}, ``{Herwig++ Physics and Manual},''
  \href{https://dx.doi.org/10.1140/epjc/s10052-008-0798-9}{{\em Eur. Phys. J.
  C} {\bfseries 58} (2008) 639--707},
  \href{https://arxiv.org/abs/0803.0883}{{\ttfamily arXiv:0803.0883 [hep-ph]}}.

\bibitem{Bellm:2015jjp}
J.~Bellm {\em et~al.}, ``{Herwig 7.0/Herwig++ 3.0 release note},''
  \href{https://dx.doi.org/10.1140/epjc/s10052-016-4018-8}{{\em Eur. Phys. J.
  C} {\bfseries 76} no.~4, (2016) 196},
  \href{https://arxiv.org/abs/1512.01178}{{\ttfamily arXiv:1512.01178
  [hep-ph]}}.

\bibitem{Bellm:2017bvx}
J.~Bellm {\em et~al.}, ``{Herwig 7.1 Release Note},''
  \href{https://arxiv.org/abs/1705.06919}{{\ttfamily arXiv:1705.06919
  [hep-ph]}}.

\bibitem{Czakon:2017wor}
M.~Czakon, D.~Heymes, A.~Mitov, D.~Pagani, I.~Tsinikos, and M.~Zaro,
  ``{Top-pair production at the LHC through NNLO QCD and NLO EW},''
  \href{https://dx.doi.org/10.1007/JHEP10(2017)186}{{\em JHEP} {\bfseries 10}
  (2017) 186}, \href{https://arxiv.org/abs/1705.04105}{{\ttfamily
  arXiv:1705.04105 [hep-ph]}}.

\bibitem{Behring:2019iiv}
A.~Behring, M.~Czakon, A.~Mitov, A.~S. Papanastasiou, and R.~Poncelet,
  ``{Higher order corrections to spin correlations in top quark pair production
  at the LHC},'' \href{https://dx.doi.org/10.1103/PhysRevLett.123.082001}{{\em
  Phys. Rev. Lett.} {\bfseries 123} no.~8, (2019) 082001},
  \href{https://arxiv.org/abs/1901.05407}{{\ttfamily arXiv:1901.05407
  [hep-ph]}}.

\bibitem{Czakon:2020qbd}
M.~Czakon, A.~Mitov, and R.~Poncelet, ``{NNLO QCD corrections to leptonic
  observables in top-quark pair production and decay},''
  \href{https://dx.doi.org/10.1007/JHEP05(2021)212}{{\em JHEP} {\bfseries 05}
  (2021) 212}, \href{https://arxiv.org/abs/2008.11133}{{\ttfamily
  arXiv:2008.11133 [hep-ph]}}.

\bibitem{Rohatgi2022}
A.~Rohatgi, ``Webplotdigitizer: Version 4.6.'' 2022.
\newblock \url{https://automeris.io/WebPlotDigitizer}.

\bibitem{deFavereau:2013fsa}
{\bfseries DELPHES 3} Collaboration, J.~de~Favereau, C.~Delaere, P.~Demin,
  A.~Giammanco, V.~Lema\^\i{}tre, A.~Mertens, and M.~Selvaggi, ``{DELPHES 3, A
  modular framework for fast simulation of a generic collider experiment},''
  \href{https://dx.doi.org/10.1007/JHEP02(2014)057}{{\em JHEP} {\bfseries 02}
  (2014) 057}, \href{https://arxiv.org/abs/1307.6346}{{\ttfamily
  arXiv:1307.6346 [hep-ex]}}.

\bibitem{LHCHiggsCrossSectionWorkingGroup:2016ypw}
{\bfseries LHC Higgs Cross Section Working Group} Collaboration, D.~de~Florian
  {\em et~al.}, ``{Handbook of LHC Higgs Cross Sections: 4. Deciphering the
  Nature of the Higgs Sector},''
  \href{https://arxiv.org/abs/1610.07922}{{\ttfamily arXiv:1610.07922
  [hep-ph]}}.

\bibitem{Biekotter:2023oen}
T.~Biek\"otter, S.~Heinemeyer, and G.~Weiglein, ``{The 95.4 GeV di-photon
  excess at ATLAS and CMS},''
  \href{https://arxiv.org/abs/2306.03889}{{\ttfamily arXiv:2306.03889
  [hep-ph]}}.

\bibitem{Ashanujjaman:2024lnr}
S.~Ashanujjaman, S.~Banik, G.~Coloretti, A.~Crivellin, S.~P. Maharathy, and
  B.~Mellado, ``{Anatomy of the Real Higgs Triplet Model},''
  \href{https://arxiv.org/abs/2411.18618}{{\ttfamily arXiv:2411.18618
  [hep-ph]}}.

\bibitem{Coloretti:2023yyq}
G.~Coloretti, A.~Crivellin, and B.~Mellado, ``{Combined Explanation of LHC
  Multi-Lepton, Di-Photon and Top-Quark Excesses},''
  \href{https://dx.doi.org/10.1103/PhysRevD.110.073001}{{\em Phys. Rev. D}
  {\bfseries 110} no.~7, (2024) 073001},
  \href{https://arxiv.org/abs/2312.17314}{{\ttfamily arXiv:2312.17314
  [hep-ph]}}.

\end{thebibliography}\endgroup

\end{document}